# Practical, scalable alternative session encryption using one-time pads

Marc Abel
December 15, 2012

Wright State University
Fairborn, Ohio
abel.12@wright.edu

## 1) Summary

When I was smaller, a five megabyte fixed disk cost $5,000, a 300 bps modem cost hundreds of dollars, and communication links were intercepted by attaching devices to the target subscriber's local loop. From then to now there have been three great implosions: the cost of storage, the cost of bandwidth, and the cost of surveillance. The wake of the first two implosions sheared away most obstacles to using one-time pads to encrypt data in flight, and the final imposition — I mean implosion — now makes consideration of one-time pads a practical necessity.

So far as assurance of confidentiality is concerned, today's block ciphers and public key cryptosystems flunk the exam. I don't know how to recover an AES key or compute the discrete logarithm of an elliptic curve element, but there exists no proof that another cannot. Moreover, encrypted communications can be recorded and stored for later attack by algorithms and devices yet to be discovered. Equally concerning is that when a significant "break" is discovered for solving either puzzle, the safety of an entire planet's communication and data dissolves like instant pudding.

The world is unready to ingest so much pudding. We need balance in our cryptographic diet, and we need that balance now. In this paper, I discredit many myths concerning one-time pads, discuss practical steps to address perceived shortcomings, and shatter the notion that secure generation, distribution, and use of mammoth cryptographic keys cannot be practiced in every home, church, school, and business. I also discuss my own implementations, their capabilities and track record, and where they should lead.

## 2) Background

Interest in one-time pads (OTPs) is not new, even for me. This research marks the third time in fifteen years that I have devoted 100% of my professional time for some weeks or months to advancing their practicality and adoption. Despite my best effort and good intentions, I have not done nearly enough, and my safety as well as the safety of our world both demand that I think on a larger scale than months and weeks. It turns out that one-time pads have a 130 year history that I can trace.

American Civil War veteran Frank Miller invented one-time pads in 1882 for securing telegraphy. No pun intended, but frankly anyone who tells you that OTPs can't be distributed conveniently is a wuss. Miller would have told you to walk, ride a horse, hire a stage, go by rail if you were well off, or send a trusted messenger if you were better off. Today I can fit 650 terabytes in a carry-on bag, although it would weigh 65 kilograms.[1] I suggest a bag that has wheels.

By 1917, Miller and his time were long forgotten, so Gilbert Vernam invented one-time pads again. Not wishing to be forgotten also, Vernam patented his discovery and teamed up with Joseph Mauborgne, a U.S. Army Captain, to build a practical machine. Although Vernam's name is heard more often now, Mauborgne was not forgotten in his time. He would become the famed "Cubic General" in command of the Army Signal Corps. Mauborgne ultimately was drilled out of the Army on account of his age — a devastating loss, as he would make violins and collect his pension for thirty more years.

By 1941, Vladimir Kotelnikov had mathematically proven that one-time pads were perfectly secure (impossible to decrypt or recover a key from ciphertext) if used correctly. As Kotelnikov wasn't permitted to share his result with the Americans, wartime cryptographer Claude Shannon had to rediscover the same properties of OTPs on his own.

---

1   Western Digital model WD20NPVT, 2 TB, 180 grams, 15 x 69.85 x 100.2 mm. Amazon.com's price as of December 9, 2012 is $61,746.75. The bag is presumed to measure 55 x 40 x 20 cm and weigh 6.5 kg.

Shannon eventually was able to publish his work in 1949. I'm glad I'm not a wartime cryptographer.

Even as digital devices became popular, one-time pads became very *un*popular. Users of early computers were as lazy as memory was expensive, so short, persistent keys became a primary design goal. Later, as people caught on that the same key shouldn't protect everything, it became the fashion to use a different and random key for every document or communication session, but encrypt all of these with a single master key in a recently invented "public key" cryptosystem. "Ciphers" that operated on small blocks came into vogue, sometimes using keys as small as 48 bits and effective key strengths as small as — in the case of Debian's OpenSSL from 2006 until 2008 — 18 bits [1].[2]

The general reputation of one-time pads is not good. Early on, the Soviets ran out of random numbers during the Cold War, forcing them to reuse their key material in a manner which American cryptanalysts were able to decode. More recently, an endless run of amateur and should-have-been-amateur programmers dreamed up methods to generate "one-time pads" from deterministic random number generators (RNGs). They ended up designing stream ciphers, which might or might not have offered some security, but they were not one-time pads, and the reputation of the latter suffered.

A general lack of nondeterministic equipment in consumers' hands perpetuated a legend that practical methods of producing truly random numbers cannot be had at reasonable cost. Perceptions were only made worse when various hobbyists and educators discovered they could use sound cards to digitize atmospheric noise; they would tune a radio to an undisclosed open frequency, and an entropy source was born. Had these hobbyists been electrical engineers, they would have realized that they could simply disconnect the antenna from their radios, as above 30 MHz receiver noise is much higher than atmospheric noise. But however implemented, when people began calculating how many years of sound card input would be needed to amass a reasonable quantity of OTP material, they again dismissed one-time pads as a novelty.

The appearance of a popular website [2] which published "true" random numbers at the snail's pace of one megabyte per day further dragged one-time pads to their present river bottom resting place. As I write, this website has delivered "1.21 trillion bits to the Internet community" — in other words less than 141 gigabytes, and it's taken 14 years to supply even that much. I can produce the same quality and quantity in less than 76 hours, and you can too if you invest $50 in some obsolete hardware and read Appendix A of this paper.[3]

Then there is the key distribution "problem", specifically that there isn't a cost-effective equivalent of Diffie-Hellman key negotiation that permits parties to generate keys over an open communication channel. There exist quantum key distribution (QKD) techniques, but they are not fast and must be carried out via a single optical hop, typically using fiber but in some instances free space. Yet the demand for the "unconditional security" of quantum cryptography beckons so many, with the result that QKD has been demonstrated for tasks as diverse as interbank communication and election result delivery [3, 4].

A scalability concern for one-time pads stems from how keys are distributed. If we were to publish a "phone book" of 10,000 OTP users, then the potential number of communicating duos is

$$\binom{10,000}{2} \ = \ 50,005,000$$

Conventional reasoning is that because we do not know which pairs may decide to communicate in the future, they cannot obtain the benefits of one-time pad encryption without predistributing more than fifty million distinct OTPs among the parties. At perhaps a terabyte apiece, this figure comes to either a lot of carry-on bags, or a locomotive pulling 48 weight-constrained "high cube" boxcars. Fortunately, I show in this paper that in most cases it isn't actually necessary to predistribute this much key material in so many combinations.

Here is a synopsis of the remainder of my paper. In Section 3, I discuss my *iid* software and protocol for storing, using, and distributing one-time pads. Section 4 presents and discusses measurement and test results. Section 5 presents a roadmap of future work. Section 6 encompasses closing thoughts, credits, and references, and Appendix A discusses rational ways of generating suitable yet gargantuan keys at reasonable cost.

---

2  This bug affected numerous computers under my charge, including specialty encryption equipment for telephone conversations.

3  I have six machines like this in storage, meaning I could finish within 13 hours if I operate them simultaneously. An hour's use of my equipment would supply the cited website for over a year.

### 3) Prototype implementation

I set out to build a one-time pad implementation I would want to use; that is, it needed to be convenient, provably secure under certain assumptions, probably secure when used in practice, scalable to thousands of users, robust, and not more complex than necessary. This work is by no means complete, but many landmarks have been seen in the distance and gradually met.

### 3.1) About notation

For ease of reading, scalar variables are designated using lowercase, italicized letters. To avoid confusion, I avoid the letter *l*. Byte (octet) strings are indicated as capital, italicized letters. I overload the symbol + to mean either scalar addition or string catenation, depending on the case of the arguments. The exclusive-or operation is written out as "xor", rather than adopt a funky symbol from the character set. Binomial coefficients are indicated by stacking *n* above *k* in parentheses. The cryptographic hash function by Professor Rivest as specified in RFC 1321 [5] is written as MD5( string ).

Examples:

> $P$ = your message
> $n$ = length of $P$ in bytes

### 3.2) Message protocol and encryption algorithm

At first blush, one wouldn't think we need to talk about "encryption algorithms" for one-time pads, because everyone says to derive the ciphertext from an exclusive-or of the plaintext and key. So let me ask ten questions:

1. How does a message recipient know which one-time pad needs to be used to decrypt the message, without having an attacker able to correlate sets of messages for traffic analysis?

2. How does a recipient know the offset to apply to the pad, without an eavesdropper being able to monitor traffic volume merely by occasional sampling?

3. How is a message protected from so-called "bit flipping" attacks, in which an interposing attacker might know or guess what information is conveyed at a particular offset and reverse the meaning of the message at that offset?

4. If a session-based protocol such as TCP is used, how is the session protected against the numerous published and

unpublished attacks that have sprung from that protocol?

5. If a connectionless protocol is used, how are packets reassembled in the correct order?

6. Also with a connectionless protocol, how are errors detected? How much assurance is needed that an assembled stream is error-free?

8. How are messages acknowledged? When an acknowledgement times out, should the missing packet's original ciphertext be reused? Can repeating a ciphertext increase an implementation's susceptibility to attacks?

9. How can endpoints guarantee that they will not encrypt with the same portion of a one-time pad, although both use the OTP simultaneously for transmission? In the event common key material encrypts two plaintexts, both plaintexts forfeit their expected assurances of secrecy.

10. How much protection is necessary against an attacker being able to examine the lengths of packets? After all, if the word *coffee* has six letters and *tea* has only three, exclusive-or with the keystream does not itself offer adequate encryption of these short utterances.

11. How much protection is sought against an eavesdropper being able to monitor the timing and volume of transmissions? Since I'm past my tenth question, I'll withdraw this one for a subsequent paper. Especially since a thorny twelfth question follows concerning an eavesdropper's ability to identify a session's endpoint locations.

I selected User Datagram Protocol (UDP) for my transport layer, because in the end our convenient Transmission Control Protocol (TCP) simply wasn't an option. People interested in intercepting or disrupting OTP-encrypted sessions are rough men and women. They play to win, and they aren't above employing disruptive methods. The plain truth is, UDP is more secure than TCP. There is no session to disrupt, no SYN or RST packets to spoof, no memory allocation as a result of invalid packets, no sequence numbers that can be inferred, and not much opportunity to monkey around. In fact, hosts listening for UDP packets on specific ports are easily firewalled such that port scanners cannot determine whether a port is listening or not, unless an application chooses to send a response.

For the purpose of discussing possible answers to my remaining questions, I present a plaintext packet encryption algorithm in Figure 1. The resulting cyphertext contains a 16 byte header, followed by a tail of zero or more bytes. For reasons explained in section 4, this algorithm is not a robust, scalable design, but it is the precise algorithm used in my current prototype. I've already mentioned the packet's tail: it's simply an exclusive-or of the plaintext with the same amount of unused key. The header is a hash-based message authentication code, or HMAC, with sixteen bytes of key material constituting its secret key.

---

take plaintext packet content $P$

set $n$ = length of $P$ in bytes

set $A$ = next 16 bytes of unused key

set $K$ = next $n$ bytes of unused key

set $T$ = $P$ xor $K$

set $H$ = MD5 ( $A$ + MD5 ( $A$ + $T$ ) )

set $C$ = $H$ + $T$

ciphertext packet is $C$ and has length $n$ + 16

acknowledgement packet is $A$ and has length 16

**Figure 1: Packet encryption algorithm**

---

In order to resist length-extension attacks and resist deficiencies in their underlying hash function, HMACs apply a hash function twice in a nested manner as depicted in Figure 1. Since HMACs are not the subject of this paper, I will not talk long about hash function selection. I chose to use two invocations of MD5. I am not right now uncomfortable with MD5 for this specific use; however, I anticipate that my successors may wish to replace it with a more recent hash function.

RFC 2104 [6] calls for inner and outer padding to be applied in the computation of HMACs, and I was not aware of this at the time I wrote my code. I won't discuss this padding here, except to say that adding four lines to my source code will bring this padding up to snuff from the RFC's perspective.

Selection of UDP packet lengths for Internet delivery is an inexact science. I clamp these packets to at most 1,416 bytes of plaintext in order to avoid limits of wired Ethernet devices. Although this worked for all my tests and should work in most networks, no specification requires the entire Internet to accept packets this large.

A message recipient selects a one-time pad for decryption by trying each until one is found which matches the HMAC

in the message. This is a magnificent, elegant approach, provided that she only has one OTP to choose from, *and* that no one might inject random packets maliciously. I discuss effects of malicious jamming in section 4, and I anticipate practical remedies in section 5.

My prototype uses two methods of maintaining offset synchronization between the sender and receiver; that is, ensuring that sender and receiver's OTPs are positioned at the same unused key material. The preferred approach is to not lose synchronization to begin with, and it works most of the time. But at other times, difficulty with firewalls, segfaults during module testing, or other confusion will cause the transmitting side to get ahead of the receiver, usually by fewer than 200 bytes. So whenever an HMAC doesn't validate for any OTPs in hand, a brute-force search through all possible offsets is made a byte at a time, up to a limit about 1,500 bytes forward. Although this approach worked well during simple tests within a small network, it also facilitated an effective denial-of-service attack against CPU resources. Sections 4 and 5 describe this vulnerability to jamming along with appropriate remedies.

The packet's header, that is, its HMAC, protects its data against not only transmission errors, but also against malicious changes such as bit-flipping attacks. As both the HMAC and its secret key are 128 bits each, the minimum of these also is 128 bits, so the probability of a compromised packet passing without detection is $2^{-128}$. I don't feel a compelled to reduce this probability further, as the residual risk and its effects are no worse than those induced by power line difficulties or "soft" RAM failures produced by cosmic rays. On the other hand, I warded off a temptation to reduce the used amount of digest and corresponding secret key to 64 bits. Although 64 bits may be enough for my personal use, I might face extra tomatoes if I advance this amount as a global standard. I've been bitten so many times by 32-bit CRC collisions that I have no difficulty becoming concerned about 64-bit collisions.

I expected at the beginning that the implementation would send many packets without waiting for earlier ones to be acknowledged, reassemble them in the correct order at the remote end, and selectively request any packets which require re-transmission. This approach turned out to be more elaborate than I had time to implement. Instead, my platform has to wait for each packet to be acknowledged before the next packet can be sent. This means that an individual session only can run at a small fraction of the line speed. On the other hand, this self-throttling also permits many sessions to coexist over a given link without mutual interference. Section 4 summarizes measurements in support of both points of view.

One effort I did finish was to ensure that the most recent transmission and acknowledgement sent and received for each session persists across runs of my test program. This ensures that sessions will resume at the same OTP offsets when either side of a link restarts.

The receiver only decrypts a packet successfully if it has the correct secret key for the HMAC; that is, the receiver is using the same OTP and offset that the sender used. Upon decryption, the receiver sends this secret key (termed $A$ in Figure 1) in the clear back to the sender as a secure acknowledgement. This 16-byte packet only contains $A$, and is not encrypted in any manner. As knowledge of $A$'s value to an attacker dissipates upon successful decryption of the packet, because the receiving OTP offset is advanced at that time, $A$ can be used as an acknowledgement packet without further protective measures. Here is one uniqueness of my protocol: identifiable portions of the one-time pad are disclosed to eavesdroppers, yet the session is still secure. There is nothing an attacker can do with $A$ by brute force to make plaintext inferences, because no plaintext is available for its calculation. Furthermore, $A$ is not of any use for replay attacks, because the most an attacker could do is *help* the receiving end ensure that its acknowledgement is delivered on a timely basis.

Unacknowledged packets are retransmitted after some time using the same ciphertext. For sessions between Michigan and Ohio, I found 0.3 seconds to be a good timeout for a packet's first retransmission attempt. For subsequent attempts of the same packet, the timeout increases to two seconds. As the typical round trip time for this link was 40 milliseconds and few packets were dropped, I generally saw few retransmitted packets. Duplicate packet handling was tested by shortening the retry timer, and duplicates were identified as such when received in order as each session records its last correct HMAC. When received out of order, duplicate packets were rejected as not matching any active OTP, making packet replay attacks of no more value to an attacker than injecting arbitrary random bytes.

I asked already if retransmitting packets and acknowledgements without any changes — without consuming additional key material — creates an advantage from an attacker's perspective. Indeed a small advantage is created; an attacker can use knowledge of retransmitted packets to measure the effectiveness of jamming. One countermeasure would be to use additional key material for retransmissions, but this would give an attacker a means of forcing key depletion by jamming, as well as require a more intelligent offset resynchronization scheme. A second alternative would be to fall back on block or stream ciphers for packet retransmission, and that places the design near a slippery cliff. I left the matter as-is, because although it's true that an attacker can learn something by benchmarking live jamming performance, similar data can be attained anyhow by installing and testing the software, or simply reading one of my papers.

The necessity that the sending and receiving end never use the same portion of an OTP is addressed by dividing each pad into pages, ensuring that the send and receive pages are not the same, and designating a side to control page turns. This mechanism is discussed in more detail in section 3.3.

As is true for several encryption protocols, it is possible for an attacker to infer some information from packet lengths. My plaintext packets have a one-byte packet type code followed by applicable data, so a chat session packet containing the text "yes" will be 16 + 1 + 3 = 20 bytes long, but a packet containing "no" will be 19 bytes long. This deficiency has been addressed by others via relatively simple approaches, but I desire a more robust approach that will require more effort. The system as I tested it leaks its plaintext lengths; at some future time, I won't permit this fault.

### 3) Storage and handling of one-time pads

High-volume generation of key material is discussed in Appendix A. High-bandwidth hardware random number generators (HRNGs) are not commodity devices, and counter-intuitively this might be for the best.[4] At any rate I consider inexpensive, scalable random number generation to be a solved problem, so I turn now to what to do with all this entropy.

Entropy I produce begins life as either a single contiguous file for small batches (like a gigabyte), or a completely filled raw disk partition for larger ones (like a terabyte). These monolithic formats work fine for production and transportation of entropy, but they are not as well-suited for use in communication systems. Instead, an *entropy installation program* is used to offload these contiguous files of key material into a more scalable disk layout that I call a *vault*. The hierarchy of filenames within a vault is depicted in Figure 2. In my system, one-time pads are broken into smaller *pages*, a doubly appropriate name because (1) they are paged into RAM on demand, and (2)

---

4   This is a chip manufacturer trust issue. For an example indicting the microcontroller embedded in your keyboard, see [12].

German OTPs of the 1920s were printed on duplicate pads of paper with a serial number on each page.

```
vault/                root directory of vault
    00000/            pad 0
        00000         page 0 of pad 0
        00001         page 1 of pad 0
        00002         page 2 of pad 0
        ...           for as many pages as needed
    00001/            pad 1
        00000         page 0 of pad 1
        00001         page 1 of pad 1
        ...           for as many pages as needed
    ...               for as many pads as needed
    pad.metadata      persistent vault information
    session.data      persistent session data
    vault.locked      hedge against segfaults
```

**Figure 2:  Hierarchy of vault filenames**

Figure 2 suggests four scalability issues with my prototype. The hypothetical maximum number of pads is capped, although trivially, at 100,000. The maximum number of pages a pad may contain is capped in the same manner. Moreover, the number of pages a pad can have must consider any filesystem limits or performance issues as to how many files can be placed within a single directory. Similar concerns stem from placing all pads in a single directory. All of these limits can be removed in a straightforward manner by permitting a deeper hierarchy of a specified depth for pads, another specified depth for pages, and a "radix" specifying how many directory entries will be created at each level. I note that it is already possible to use symbolic links to span a vault across multiple volumes.

For performance reasons, the small amount of metadata needed by each OTP is not stored in the same directory as the pad itself. Instead, this information lives in a small text file that all pads share named "pad.metadata"; a snapshot appears in Figure 3. I wanted a text format not only to dodge endianness considerations, but also to permit manual resynchronization of metadata after a mishap.

Figure 3 is nearly self-explanatory, although the reader should disregard the line for pad 00000 for a moment. After the header line (which in the ASCII file is not actually in bold print), one line per pad appears. The pad serial numbers in the left column must be unique, but the serial numbers might skip around and appear in any order.

| pad | kb/pg | pages | tx pg | rx pg | tx off | rx off |
|---|---|---|---|---|---|---|
| 00000 | 04096 | 00080 | 00000 | 00080 | 00000000 | 00000000 |
| 00001 | 00512 | 00256 | 00054 | 00053 | 00085898 | 00085114 |
| 00002 | 00512 | 00256 | 00022 | 00005 | 00113459 | 00000752 |
| 00003 | 00512 | 00256 | 00022 | 00005 | 00046155 | 00000425 |
| 00004 | 00512 | 00256 | 00022 | 00005 | 00046189 | 00000449 |
| 00005 | 00512 | 00256 | 00022 | 00005 | 00113496 | 00000412 |

**Figure 3:  pad.metadata**

Different pads do not need to have identically-sized pages or the same number of pages; these parameters appear as the `kb/pg` and `pages` columns of Figure 3. The size of each page is specified in kilobytes, must be a multiple of the storage device's block size, and cannot change within a pad.

As mentioned near the end of section 3.2, a session endpoint uses a given page for either transmission or reception, but never both. This prevents conflict with the other session endpoint's page use, and maintains the invariant that any given portion of key material is used at most once. The currently active (and therefore loaded into RAM) transmit and receive page numbers appear in the `tx pg` and `rx pg` columns. These index from zero, and a pad is said to be *out of pages* (exhausted) when a page number is greater or equal to the number of pages that pad contains.

The offset within an active page is stored in the `tx off` and `rx off` columns. This offset is zeroed each time a page gets turned.

Whenever an OTP is installed at session endpoints, the transmit and page numbers are initialized oppositely; e.g., the Union transmits on 0 and receives on 1, but the Confederacy transmits on 1 and receives on 0. Page turns are coordinated by the party with the higher transmit page number. In this example, the Union does not coordinate page turns. When the Union runs low of entropy for page 0, it sends the Confederacy an encrypted packet requesting allocation of a new transmit page. The Confederacy will reply with the next available page number, page 2 in this case. Thereafter, the Union has the higher transmit page number, since the Confederacy has not exhausted page 1, and so the Union will coordinate the next page turn.

There are small granularity losses in the paged entropy scheme. Key material is lost at the ends when pages get turned; otherwise it becomes easy to run out of entropy and be unable to continue communicating. Also, a pad that is out of pages in one direction might yet have much of its last page available in the other direction.

The layout of pad.metadata causes some implementation limits. The maximum page size is 97,656 kilobytes, as the next kilobyte would overflow the eight column page offset field. The maximum number of pages cannot exceed 99,998 because of the width of that field, which must reserve the numbers 99,998 and 99,999 to indicate exhaustion of the two loaded pages. These fields cannot be identical even when a pad is exhausted, or page numbers would be confused in the event the pad gets replenished. Further limits are imposed by the number of inodes the underlying ext3 filesystem permits per directory, lowering the maximum number of pages per pad to 31,998. In sum, the maximum pad size for the present system is 2.91 TiB. The number of pads is limited to 31,995 on account of the ext3 inode count issue and three other files plus `.` and `..` in the vault root directory. It is readily apparent how to widen these limits to support anything an underlying filesystem would be capable of supporting.

The division of OTPs into filesystem-managed pages is particularly helpful as pads are consumed and replenished. Once a section of pad is spent, its retention is not just valueless, but in fact dangerous. Once a page has been exhausted and turned, it can be deleted, reclaiming space for installation of further key material, whether for the same or a different OTP. My implementation doesn't yet delete page files to reclaim space, although their content is irretrievably removed.

### 3.4)  Physical security and one-time pads

Physical security of OTP material demands rigorous attention outside and inside the machine's case. Lucid understanding of the host OS, filesystem, disk subsystem, and network is critical. In principle, there are five classes of locations where one-time pads may be inadvertently or maliciously copied, retained, and ultimately compromised. These classes are locations in RAM, on disk, on the network, in hosted environments, and upstream within the entropy supply chain.

### 3.4.1)  Locations in RAM

Data in RAM is vulnerable not only to unauthorized users (network exploits), but also to unwanted visitors who arrive by night with screwdrivers to open the case and 1,1,1,2-tetrafluoroethane to keep the chips cold while the power is off. If your applications don't maintain a silicon grip over where sensitive material resides, your OS kernel and utilities will be your enemy's informants.

When carefully written, C and assembler programs can do a good job at following up after addresses where sensitive information has been placed. On the other hand, address-agnostic scripting languages such as Python that manage memory and check buffers on your behalf don't fare so well. Unless you wrote the script interpreter yourself or are otherwise certain of the present and future versions' safety, your scripting language cannot be trusted with key material. This drawback extends also to most system programs you may use to transfer one-time pad material, such `cp`, `dd`, and `rsync`.

Even if a system program doesn't transfer any key material, a problem may surface nonetheless. Utilities such as `md5sum`, which you might use to verify the integrity of an entropy page, will load sensitive content as they calculate, and it's not very common for them to see that the content is re-secured when they terminate. Even if `md5sum` cleans up its own memory, it still will smear everything it reads all over the kernel page cache.

Although you can't permit *any* system program to access your OTP material, an army of bumbling processes are arrayed against you. `updatedb`, for instance, will index your OTPs for searchable text, thereby treating the buffer and page caches to lethal secrets. Desktop managers and automounters peep into connected media to mount and open them on your behalf. Removal of unnecessary services, as well as careful assignment of access rights, may prevent some slip-ups. It's a jungle out there, and you need to think about the lions and hyenas.

Most operating systems enable dangerous cache functionality for performance reasons. In Linux this includes the page cache, which retains as much recently-read file content as memory will permit, and the buffer cache, which retains raw disk blocks. These caches are subject to attack long after a program which accesses one-time pads has terminated. It's important that whenever OTP files get accessed, system buffers don't.

Some security against accidents could be gained by storing all key material in an unmounted raw partition, but make sure it has an unused gap at the beginning, since automounters and format identifiers like to poke around there.

Even once you've tested and proven that you got everything right with all the caches and system utilities, you still didn't. Your hard disk has its own separate RAM cache, and a wild goose chase around the disk surface might help you lose a few of your stowaways.

If you combine OTP functions with authentication mechanisms or further key material, your implementation may have non-OTP secrets which are RAM-resident with all the aforementioned vulnerabilities. If power disruption is a potential attack vector, consider using extra RAM to store the secret. For example, an MD5 digest of a 64 KiB buffer could be used as a 128-bit key, potentially increasing the likelihood that the key will be unrecoverable after the power is restored.[5]

Conformity of the present system

My implementation is very careful around RAM. I *do* use a Python installation script to set up OTP vaults, but that script only handles metadata. Transfer of entropy is accomplished via a C subprogram that is initiated by the script; this program allocates its own buffers and handles open files with the O_DIRECT flag set. My main OTP communication program is entirely in C and uses the same buffer precautions. Test results show that I am clean concerning the Linux page cache which buffers files, but I am less certain whether the buffer cache which manages raw blocks is being bypassed correctly.

Because the I/O for the installation software does not use the kernel's buffers, it runs noticeably slower than one might otherwise anticipate. Although I took no measurements to evaluate disk performance, disk latency may need to be worked around when OTPs are used in high-bandwidth applications.

### 3.4.2) Locations on disk

One of the first places to look for sensitive information on a confiscated machine is in swap space on disk, whether the space is managed as files or partitions. I quit using swap space in the late 1990s as soon as I had a machine with enough RAM to hold together, and I never looked back. In the swap sense my implementation is secure, but only because I have all swapping disabled. I have yet to add or test POSIX's `mlock` or `mlockall` system call to prevent writing pages to disk in the event swapping is active. For those who use encrypted swap space: that precaution alone is not strong enough, because the swap space's encryption key is not an OTP in its own right.

It's important to consider the behavior of the filesystem, along with how your system will be administered, to ensure that it is written to the disk in only one place. Ordinarily a filesystem will not relocate information which has been committed to a disk's surface, but exceptions are common. Defragmentation programs, for instance, will physically move what has been written to their own notion of an "optimal" placement. If the location from where OTP content was moved is not overwritten with new information, the OTP is compromised, as the cryptographic system has no means of finding the consumed entropy to erase.

It may be possible to overwrite ("wipe") a volume's free space after problematic operations such as defragmentation; however, such mechanisms are not consistently reliable. Tools which simply write to a volume until they no longer can, easily miss on the order of 5% (which can be 200 GB on a large drive) because space has been reserved for the superuser. The naive remedy is to be superuser when wiping free space, a better remedy is to know what you are doing, and the best remedy is to preclude this need.

Another potential problem is when a filesystem writes data (not just metadata) to a journal before committing the information to its eventual home. The ext3 and ext4 filesystems behave fine if mounted with its default option `data=ordered`, or with `data=writeback`, but if a volume is mounted with `data=journal`, all data will be written to two locations of the disk. `data=journal` is not a good choice for volumes containing OTPs.

Many hard disks on the market today offer internal encryption via AES; these disks can be completely "erased" in milliseconds by destroying internal key material. Such erasures are at most as strong as AES itself, and there are many prequalifications (a functioning drive, software installed, credentials available, etc.) for such to succeed.

Caution is needed when removing files containing key material, particularly if a well-validated process is not being employed. During testing on a machine in Australia, I inadvertently removed an intermediate file that had been used to set up key material by hand. This caused a dilemma: downstream copies had already been used to transfer data "securely" between Sydney and another machine in Ireland, so I had to wipe the intermediate file. As the file that needed erasure was no longer available, I wound up scrubbing the entire free space of the volume.

Full destruction of key material transferred via removable media must be guaranteed.  This is problematic for solid-state media that support write leveling, because their blocks are rotated among a pool of spares.  Moreover, if entropy is not installed directly to removable media but staged on an device internal such as a fixed disk, the staging area must be purged with caution.

---

[5] I have sought data on RAM decay rates without success.

Transfer to, from, and within *any* media immediately raises the RAM data remanence concerns of section 3.4.1.  It is not safe to `cp -ar src/ dest/` one's way around with key material.  Special programs need to be employed similar to the installation utility I wrote for this project.

Another media handling hazard is that generated entropy could be used more or less than exactly two times before destruction.  That's right, *two* times, although we still call it a one-time pad:  one time at the sending end, and one time at the receiving end.  Once key material has been installed for a specific page of a specific pad, it must not be reused elsewhere.  Although this sounds like a simple invariant to maintain, it needs to be maintained, sometimes a program fails, a media error or full volume occurs, or a user becomes inattentive or confused.  Special utilities should perform rudimentary checks for duplicate pages, as well as for pages that appear to be non-random.  As with any other processes that touch OTPs, these utilities also need to be safe from a remanence-in-RAM perspective.

Although good practice demands that essential data be backed up, one-time pads are an important exception.  Never back up a one-time pad.  Instead, have a *backup one-time pad* ready for connections where high availability is needed.  Just be certain your backup is on a separate, tamper-evident medium.  Be advised that some organizations have staff and resources devoted to circumventing tamper-evident seals.  I like to use paper currency in tamper detection schemes; it's hard to get replacements with matching serial numbers.  Just don't run afoul of any laws in your jurisdiction.

Any OTP material that remains on disk after use is a vulnerability.  The correct handling sequence is to load the key material into RAM, then obliterate the key from the disk, perform the encryption, destroy the copy remaining in RAM, and only thereafter transmit the encrypted data over the network.

Once you have done everything else correctly, it is perfectly okay and quite appropriate to run your key material through a block cipher prior to use.  The other end of the session must do the same operation using the same symmetric key, which may be distributed either by QKD or by a less exotic but guarded method.  This at best is secondary cryptography to employ in the event of a complete pad compromise.  Section 3.4.5 touches on some devastating attack mechanisms against pad integrity.

Conformity of the present system

My system is designed to overcome many disk-related key material vulnerabilities.  It turns out that I was a little inattentive with respect to swapping, and left it enabled inadvertently at the Michigan endpoint.  This underscores the need to address configuration-related vulnerabilities in the source code proper as well as in the configuration process.

Adding `mlock` functionality to prevent key material from being swapped to disk will require special consideration at installation time, because many systems by default place stringent limits on how much memory a non-privileged process may lock.[6]  For security reasons, the program should decline to run if it cannot lock enough RAM.

None of the systems I tested with have indexing programs such as `updated` enabled, nor do any perform defragmentation.

Two machines I used for testing run `ext3` for their filesystem; the four others run `ext4`.  As each used the `data=ordered` mount option by default, there was not an issue with journal file data remanence.  What should be added is testing for hazardous mount options for all programs that handle key material.

I wrote a utility that carefully transfers key material from a large contiguous file into the vault hierarchy of Figure 2, but it only considers one of the communication endpoints.  Although the program is designed to write two copies of an OTP, with one being to fixed media and one to removable media, nothing has been written to safely transfer the removable media's content into its eventual home.  What is written so far handles the all pages with code written in C, performs O_DIRECT transfers via its own buffers, and methodically marks and overwrites the input file as it is consumed.

Because key material transfer is a special process,[7] a tool was written to produce a tracer file that was substituted for source entropy at the time the OTP vault setup program was tested.  Rather than containing random bits, the tracer file was marked with human-legible byte offsets which were tracked through the vault file hierarchy to ensure distribution into the correct pages.

---

6  The default limit on the Ohio hub machine is 64 KiB per process, even though it has 4 GiB of RAM installed.

7  Special processes are described in Section 3.4.5.

### 3.4.3) Network-accessible locations

The moment you copy a one-time pad across a network, the entire pad has been irreparably compromised. Three exceptions to this rule are if (1) the one-time pad will never be used to protect sensitive information, (2) the network is so small that you control it absolutely, or (3) you encrypted the one-time pad with a second one-time pad as you copied the first across the network.

I claimed the first exception for my testing, as I had machines set up in Ohio, Michigan, Virginia, Oregon, Ireland, and Australia. I had physical access to and control of the Ohio machine only; I did not actually travel to the other locations to conduct any testing. As expected, no sensitive information was ever or will ever be protected with the key material I used for testing.

The second exception applies if you have directly interconnected two machines on your desk to transfer sensitive material, and the possibility of compromising emanation attacks has been duly considered and somehow addressed.

The third exception is of enormous value, because you can use it to enable an OTP star network to function as it were an OTP ad-hoc network.[8] In other words, if you distribute $n$ OTPs between $n$ endpoints and a common hub, the hub can transmit new OTPs to any combination of endpoint pairs. It solves the freight train problem of Section 2 with just 10,000 OTPs pre-distributed instead of 50,005,000.[9] You must be cautious defending your hub, as it is a high profile asset from an attacker's perspective.

One paradox of communicating using one-time pads via a network is that the network must never touch the OTPs. Not only must you never transmit key material through a network, but you also can't enable access to the key material. All conventionally encrypted network protocols are off-limits for you, because their keys are all weaker than the OTPs you are safeguarding. If block ciphers and public key schemes are inadequate to safeguard your important transmissions, they also are inadequate for keeping script kiddies out of the OTP material which secures these same transmissions.

I'll say this another way: you can't run `sshd` or any other remote access program on a machine that has an OTP installed.[10] I advise removing sshd and every other network service you can think of from any OTP hub server always, and from the endpoints if possible.[11] Consider rebuilding your kernel without TCP support at all, as this may evade a backdoor or malware you would otherwise miss. Set up firewalls — not just incoming firewalls — and don't install any more "security updates". Buy a USB GPS receiver for $24 to keep your system clock accurate, because you'll no longer be able to set it using the Internet.

If software is available to login to your hub via one-time pad, you may install this if you absolutely must. Otherwise, manage your hub at the device itself, because if you manage an OTP hub from another OTP endpoint, you double the number of places it can be attacked from. Your command link may be exploited to install rogue software, such as that TCP stack you carefully removed, to introduce additional weaknesses.

OTP hubs are also sensitive to attacks against their stored entropy and HRNG implementations. If an attacker can replace a hub's undistributed, presumably secret entropy with deterministic material generated by seeds known to the attacker, every machines using key material from the hub will be compromised. As discussed in Section 3.4.2, the network should include an additional block cipher as a failsafe.

<u>Conformity of the present system</u>

As the hub I tested ran on a Linux desktop without any remote access enabled, it was moderately secure in terms of network reachability. On the other hand, the endpoints used for testing were network accessible to any attacker capable of circumventing OpenSSL [7].

### 3.4.4) Hosted locations

There are some setups where OTPs might not help you. The classic example is if you lease a server in a datacenter you have never set foot inside. You need absolute physical control of your OTP systems. Likewise, using a one-time pad within someone else's virtual machine monitor (VMM) is deranged.

Every rule has an exception. If you are conducting tests for an OTP research project, you might rent a cheap VPS as I did for five of the six systems I used.

---

8   Figure 7 may help the reader visualize an OTP star network.
9   The 48 boxcars can be replaced now with 16 carry-on bags.

10  If you suspect I have an exception in mind, it's in Section 3.4.4.
11  An `ssh` client without incoming access is safe.

You need absolute logical control in addition to absolute physical control of your systems. OTPs aren't safe on machines that run closed-source operating systems, including so-called "shared source" systems. What you can do instead is proxy these machines through an intermediate OTP firewall that uses an open-source operating system.

When you run an open-source operating system, consider the origin of the binary distribution you installed. No law protects you from any government's ownership or control of software companies with names you might trust. Also consider that machines used solely for as OTP firewalls might not need much of an operating system. Less is more.

There are good reasons to use virtual machine monitors in OTP devices. One framework may use a VMM to create a separate sandbox to hold key material and do the necessary cryptography. Separate virtual machines within the same VMM might have full network stacks, enable ssh logins, get the time from the Internet, and other extras.

Another framework might use a small VMM as thin network and block device drivers for a client kernel that does nothing but OTP operations. The benefit here is that the VMM code base can be small enough to be relatively free of exploits, as well as small enough to audit its source code in-house. With device drivers of this VPS already in place, the client VM's OTP-functioning "kernel" and UDP stack can be implemented with just a few thousand lines of code.

These are only the foothills of the trusted computing mountain range; you still need to consider your BIOS, your CPU, and a few other VLSI devices before trusting that your communication is private. I nevertheless place strong emphasis on the likelihood of concealed defects in the operating system, whether or not it is open-source. Influence over operating system distributions is the most cost-effective cyberattack that a taxpayer-supported adversary can mount.[12]

Conformity of the present system

As described earlier, I did employ virtual private servers running within VMMs running on equipment under other people's control for this research. All operating system code used was from canned binary distributions acquired

---

from other parties and verified via MD5 and/or public key cryptography. In short, I would have handled everything differently had I needed and had time to implement greater security.

### 3.4.5) Locations upstream in the entropy supply

Up to this point, this paper assumes the availability of an untainted, protected supply of nondeterministically generated, independent and identically distributed blocks of ones and zeros. Perhaps you generate these blocks at the time of installation, or perhaps they are already on media. Perhaps you know their source. Perhaps they came from an OTP hub operator. But unless you control the supply chain, there is no way to examine the medium or content to assure that this content is (1) random, or (2) secret.

I devoted several months studying OTP supply chain issues and designing defensible mechanisms for generating and distributing random numbers in high volume. The culmination of that work is a working supply chain designed to conform to both ISO 27001 [8], which specifies requirements of information security management systems, and ISO 9001 [9], which does the same for quality management systems. A portion of this supply chain is a set of fast hardware RNGs along with software to control it, but there is much more involved.

Production of keys for cryptography is what ISO 9001 considers a "special process"; that is, a process after which conformance of its end product cannot be proven by inspecting the end product. A correctly functioning HRNG is as likely to output an exact duplicate of this paper as it is any other fixed output of the same length, even though most other output combinations would be thought to be "more random". For this reason, an entropy supply chain for OTPs requires near-impossible scrutiny and continuous validation and verification.

Conformity of the present system

A risk assessment I conducted against my OTP supply chain identified 185 distinct vulnerabilities. Note 185 cases of the same vulnerability, such as 185 files placed on a single medium, but 185 independent risks requiring assessment and treatment. The supply chain itself was very simple; all that was being manufactured were random numbers written onto media. The production equipment operated very transparently, too. It used no fixed media and was never attached to any network. Network support was compiled out of the operating system entirely. Yet with all those unknowns removed, 185 attack vectors remain to cope with.

   

Although I at one time had several terabytes of OTP keys on-hand, my work for this paper used only 1 GiB of what I have on-hand. I generated this entropy on June 18, 2009 using version 1.01 of my unpublished *Eternity* OTP production software and gave it batch number K8CLZN for tracking purposes. I also recorded which machine produced the material so that I could recall the batch in the event hidden defects with the hardware were identified later. That machine's name[13] was *Day 1*.

I was in the room from the time *Day 1* booted up until both copies bore tamper seals with recorded serial numbers. When I was ready to use the batch for this project on November 29, 2012, I verified these serial numbers prior to inspecting and removing the tamper seals. Prior to installing the key material, its MD5 digest was checked,[14] and after installation the partitions on both flash devices were overwritten from `/dev/urandom`.[15]

In this discussion about the entropy supply chain, I've done no more than to allude to the magnitude and complexity of the problem. There remains much to be written in other papers about how this problem may be managed.

### 3.5 The *iid* one-time pad program

The program I wrote for the bulk of my testing comprises 4,000 lines of C. It provides a repository for one-time pads in the manner shown in Figures 2 and 3, code for encrypting and decrypting plaintext, transmitting and receiving UDP packets to and from multiple sites simultaneously via IPv4, acknowledging and resending packets as needed, obtaining configuration settings, diagnostics, logging, and a run loop. A split-screen user interface using ECMA-48 escape sequences[16] is provided.

My program name *iid* derives from the statistical term *independent and identically distributed*, because the UDP layer of its communication protocol is exactly that. This is a little irregular, as a majority of encrypted protocols send some non-encrypted metadata. For example, the sixth byte of every Transport Layer Security [10] record will tell any attacker which version of TLS she is up against.

---

13 The six production systems were named after the six days in the Genesis creation narrative.

14 This, unfortunately, triggered the RAM data remanence problem described earlier. To clean that up, I booted the computer into a lengthy RAM test which overwrote all contents. Future pad installation tools will provide safe digest functions.

15 Writing zeros might be a bad idea, in case a device appears in the future that treats the zeros as a sparse file and compresses them.

16 Although these are commonly known as ANSI escape sequences, it turns out that ANSI withdrew the standard fifteen years ago.

As it would be simplest to describe *iid*'s function in light of its inputs, sample configuration files appear as Figures 4 and 5. The first of these applies to OTP hubs that distribute new key material to endpoints, and the other to vanilla endpoints.

| | |
|---|---|
| ; Server configuration for Ohio OTP hub | |
| Server | ; this is the hub |
| ListenOn 49494 | ; port number (not IANA-assigned) |
| Vault "/var/otp" | ; directory for metadata and OTPs |

**Figure 4: Server configuration file**

Although the hub has more work to do than an end-user machine, its configuration file is smaller. It needs but to know that it's running as a server, where to look for its files, and which port to listen on. Provided that UDP traffic can reach that port and the pads are installed,[17] it's ready to go. You don't have to tell the server anything about the client locations, because they announce their addresses and port numbers transparently within the UDP protocol.

Client addresses are only recorded when an incoming packet is successfully decrypted. Although an interposer can exploit this mechanism to mess up a hub's understanding of clients' locations, this adds minimal exposure given that same interposer's ability to block all traffic in the first place. Consideration was given to having the endpoints encrypt their locations into the ciphertext, but endpoints will often be in subnetworks where they do not know their externally reachable IP addresses and port numbers. Thus I rely on the UDP packet headers, even though they are neither encrypted nor authenticated.

Client configuration is only slightly more complex; you have to tell it where the server is. You also have to tell the client who it is, in the sense of which OTP to use to communicate with the server.

If you're going to use the client to transfer files, you must designate a directory for them to be delivered to. *iid* will add suffixes to incoming files as required to address name collisions. There are no further security options for file transfers, because only the local user of the program (who presumably has shell access) can initiate an outgoing transfer. If no receiving directory is provided for incoming files, then incoming transfers are disabled. No space limit is imposed for incoming file storage; in fact, the space issue asymptotically takes care of itself if the `RxFiles` and `Vault` directories live on the same volume.

---

17 Figures 2 and 3.

```
; Client configuration for Oregon VPS

User 3                        ; pad # to communicate w/ server
ListenOn 49494                ; port number (not IANA-assigned)

ServerAddr "hub.abel4.us"     ; Marc's machine at home
ServerPort 49494              ; port number of remote server

Vault "/var/otp"              ; directory for metadata and OTPs
RxFiles "/tmp/rx"             ; directory for received files
```

**Figure 5:  Client configuration file**

When *iid* starts, if there were any sessions from the previous run with pending unacknowledged packets, *iid* resends these packets to their most recently known network destination.  Note there is at most one unacknowledged packet per session.  Because a session "belongs to" an OTP, not to a network endpoint, it is possible any number of sessions to be active between a given pair of endpoints, even though the same port is used.  *iid* does all its work with a single thread and single socket, calling `select` to multiplex input from the network and user's keyboard.

A little handshaking is helpful, but not strictly necessary, to alert users to connectivity problems, reduce opportunities for the endpoint OTP offsets lose synchronization, and preclude entry of sensitive plaintext if an endpoint cannot transmit in the first place.  Firstly, if pending acknowledgements (ACKs) are not cleared for any sessions, these sessions will block until connectivity can be re-established.  This includes ACKs which were being waited for at the time the previous run of *iid* terminated, since *iid* persists its session state between runs.

Secondly, a *connectivity probe* packet is sent to the remote endpoint, and if it is acknowledged, a *round trip established* (RTE) packet informs the remote system that packets have been sent in both directions successfully.  These actions cause both sessions to enter a *connected* state where they are ready to convey traffic.

A session which does not become connected has little to go on for troubleshooting, because it will not receive any kind of a reply.  This is by design.  For example, there is no provision in the protocol to say "you aren't a valid user", "you're at the wrong offset in the key", "I'm out of entropy right now", or "I know that you can't find your OTP, but I'm answering your test packet".  Eloquent silence answers all mistakes.

If a connectivity probe is not answered, it will be resent periodically.  If the program shuts down and is restarted, the same unacknowledged probe is sent again.  Note that unacknowledged packets are retained only as ciphertext; by the time a transmitting endpoint has finished encrypting a message, it has neither the message nor the key material available to read it again.[18]

Figure 6 shows the *iid* user interface running inside a terminal window.  The design is similar to any of many chat programs; you type into the blue bottom area, and when you press Enter it gets sent across the network to someone else who is running the program.  Both sides of the chitchat scroll through the black top area.  The locally typed text is displayed as blue (because it "came from the blue area") to distinguish it from what the other side typed.  Of course, *iid* is distinct from other instant messaging software in that the conversation is encrypted with a one-time pad.[19]

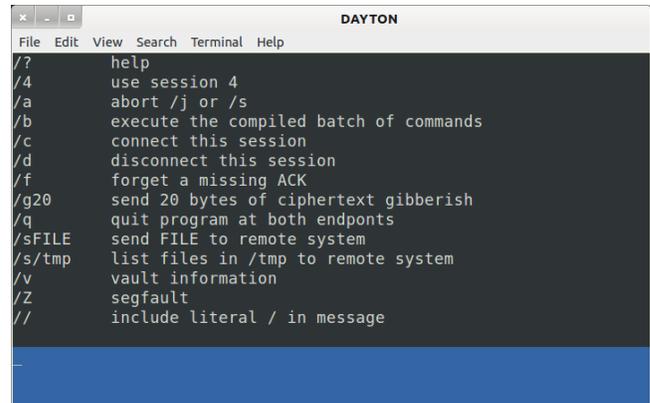

**Figure 6.  *iid* main screen and help message**

Commands to *iid* are preceded with a slash in order to distinguish them from conversational text.  If you want your text to begin with a slash, you may type two in a row.  The local user in Figure 6 has just typed the "help" command, `/?`.  The remaining commands are discussed below, although not in alphabetical order as shown in the figure.

The next command after `/?` (help) is the command to switch which OTP you are using, and thereby which client you are talking to.  You do this by typing the pad number

---

18 Persons astute to data remanence issues may observe that the program's terminal window is scrolling the chat dialog.
19 One exception is [13], an OTP plugin for another chat program.

                    

after the slash; e.g., `/5` addresses the session using pad 5. (This pad appears in Figure 3 as 00005.) Any subsequent typing and/or commands that pertain to a specific session will go to that session. Note that if that session is blocked for an ACK or connectivity probe, certain commands as well as chat text will be rejected and an explanation given.

For testing purposes, a `/d` (disconnect) command causes the session to forget that connectivity has been established. This command also goes across the network and puts the other endpoint in the same state. The session stays disconnected until the user requests a new connectivity probe using the `/c` (connect) command.

If a session is blocked waiting for a packet to be acknowledged, it is possible to force it to unblock; that is, simply forget about the ACK. That command is `/f`, and it was helpful for earlier module testing. In most cases, the remote endpoint will need to resynchronize the receiving page's offset once connectivity is regained.

*iid*'s normal exit is via Ctrl-C. The `/q` command provides a stronger quit; it will cause the program to close at both endpoints of an active connection without the user needing to take either hand off the keyboard. An even stronger local quit is available with `/Z`, which must be capitalized; the program will segfault intentionally.

Care was taken to ensure that *iid* has a single exit point.[20] This is important because of the amount and fragility of sensitive data under the program's care. Almost any system call the program uses is subject to fail, particularly during initialization, and it's important that the frequent task of starting the program never in itself reduces security or increases any person's workload.

Not all program crashes are the fault of the program, operating system, or configuration. Equipment failures, sometimes from causes external to the earth's atmosphere, cause programs to crash at times. *iid* does a good job addresses several needs which surface in the event it does crash.

First, we need to consider the fate of the active transmit and receive pages. They aren't on disk anymore, because the files were overwritten with `/dev/urandom` when they were loaded into RAM. That's good. Nor is any consumed portion of these pages in RAM still, because all consumed areas were overwritten at the time of consumption. Also good. But any unconsumed portion of a page is a problem, because this sensitive information continues to exist in memory that has been reclaimed by the operating system. That's not good. No one can tell what a kernel might do with secret information it isn't supposed to have.

Computers have an uncanny knack for doing exactly the worst possible thing. When mine sees *iid* crash, it produces — a core dump file! No not only has secret key material leaked into a memory page that no one knows which process may get next, but it's also on the disk surface, an unknown number of I/O buffers, and by now probably loaded into some debugger (with its own separate leaks) as I work to assess the place where the crash occurred. We now consider whether or not all this disclosure presents a security issue.

The crash that just happened left the OTP pages out of synchronization at the connection endpoints. It's not that the offsets are out of synchronization; they might be or not, but the pages themselves don't match at all, since one side went south with bogus entropy from /dev/urandom in the transmit and receive page files. That's good, because we don't want to recover these pages anyway. They've been leaked all over Ohio by now. Now if you look back at the last file in the Figure 2 vault hierarchy, you'll see a file named `vault.locked`. It's created when *iid* starts, and it's removed when *iid* terminates normally. But after a crash, `vault.locked` is still there, and on the next startup *iid* will find it, issue a short apology, and terminate.

The user must manually recover both endpoints from the segfault. The vault.locked file must be removed, and pad.metadata must be hand-edited at both endpoints to turn to the next two unspent transmit and receive pages and zero their offsets. Note that the transmit and receive page numbers and the two endpoints must be backwards from each other. If the system that crashed was an OTP hub with a large number of pads configured, they all must be fixed in the hub's `session.metadata` — as well as on every remote machine.

I spent so much time turning pages on six machines after various abnormal exits of *iid*, that I added a configuration option to suppress the program's automatic destruction of segfault-invalidated key material until my code was more stable. The name of this option, which is not shown in Figure 4 or 5, was HaveMercy. *iid* still incorporates this diagnostic feature, although it is removed today from all configurations.

A `/v` command displays vault information, which in essence shows what `pad.metadata` would contain if it

20 Except for the `/Z` command for testing miserable situations.

Speak softly, and carry a big key.

were to be written at that instant. An extra column is added to clarify whether or not the local end controls OTP page turns, although the same information is available by examining the page numbers. Output from `/v` is visible in Figures 9 and 10.

To test OTP page turn code, I needed a means of consuming large amounts of key material. This command is `/g`, and it sends ciphertext gibberish to the remote endpoint. As many packets as necessary are sent, and if needed the two final packets are rebalanced in length to get the total "byte on". Because the receiving side checks and acknowledges every packet, there is no reason to display the received data. Once the last ACK has been received, the elapsed time that the transmission took is shown.

The `/s` command sends a local file to a remote endpoint. If the file turns out to be a directory, it is listed for the remote user. No metadata such as modification time or permissions are transferred. In deference to the privacy of others who might test this software with me, there is no corresponding command to list directories on or retrieve files from a remote machine.

An `/a` command aborts transmission of gibberish; it also can be used at either end to discontinue an in-progress file transfer.

Finally, there is a `/b` command that executes a compiled-in script of other commands. Its purpose is to give the tester a way to start data transfers on more than one session at a time for benchmarking purposes.

It should be kept in mind that *iid* is simply a framework for one-time pad protocol, security, and usability research. It's not an instant messaging program. For one thing I don't IM, and for another, robust support for keys as large as 2.91 TiB could be overkill for a chat program.

### 4) Results of testing and measurement

Qualitative results, speed measurements, and the effect of an attacker injecting unwanted packets are presented in this section.

### 4.1) Qualitative outcomes

Sometime close to when I finish writing a large program, it usually needs rewritten from the ground up. I am happy to have found time for that rewrite, but *iid* is not completely stable yet. When I started taking measurements in earnest at endpoints located on different continents, the endpoints needed coaxing that bordered on cardiopulmonary resuscitation to get the various network addresses, OTPs, and encryption offsets working together. Everything tends to keep working in subsequent runs once motion has been established, but the work in getting to that point tells me that my platform isn't *usable* today. And while I won't offer a numeric usability measurement, I do see this as an experimental result.

On some machines, I had to manually turn the transmit and receive OTP pages twice to establish communication. There is a good reason I might have to turn the pages once in the event of an abnormal termination, but *twice* indicates that confusion reigned at some point. Again, this is a code stability and usability result.

There was a specific functionality that I wanted *iid* to provide above all others: I wanted the program to distribute new OTP key material across a network securely, because this is an important theoretical and practical result. You don't have to predistribute $n^2$ keys to enable secure communication in an *n*-user network. *n* keys is sufficient, if you have a secure facilitating hub which can distribute new keys on demand. This aspect of my project was a near-failure, because the functionality wasn't ready by press time. I felt very strongly that reducing this method to practice was vital, even if I knew from the outset that it would work.

I had thought it was a little vain to build a file transfer capability into *iid*, as it wasn't essential to demonstrate an ability to exchange data. Looking back, I can see how supporting file transfers helped unify and clarify the handling of multi-packet transmissions. An unforeseen consequence of this "vanity feature" was that the day was saved concerning key distribution.

Although *iid* did not natively support distribution of new key material across an OTP-secured link, I was able to use *iid*'s file send capability to manually send new OTPs across the line securely, and then manually install these new keys at their endpoints. Having done exactly this, I have completely reduced my key distribution scheme to practice, although I concede that I did not automate it well.

My test setup uses six hosts on three continents. My desktop machine in Ohio serves as the OTP hub; its speed is capped by a 768 kbit/s DSL uplink rate. A wireless LAN adds slightly more than a millisecond to the round trip time of all packets exchanged with the Ohio host. A small VPS with 128 MiB of RAM is located in Michigan. The other four machines are 613 MiB VPS instances located in Virginia, Oregon, Ireland, and Australia. Figure 7 shows a picture of the network topology with 128 MiB keys distributed between the Ohio hub and the five otherwise unattached endpoints.

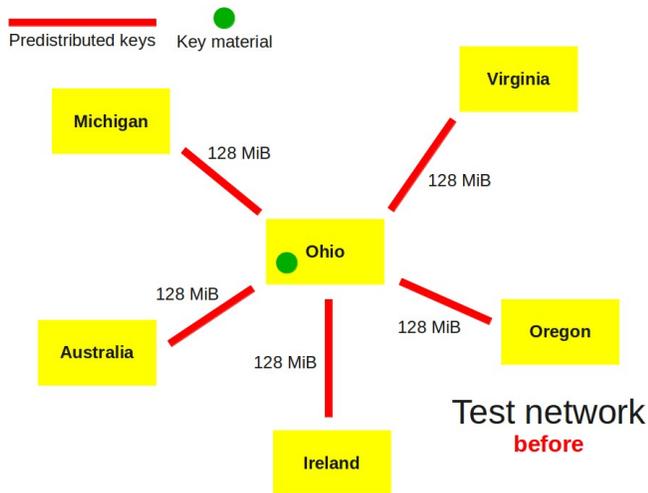

**Figure 7: Test network before conveying OTP material**

Pad number 00000 is reserved for unused key material retained by OTP hubs for future distributions. This pad is shown in the `session.metadata` listing of Figure 3, and as a green dot near the center of Figure 7. One difference between this pad and the others in Figure 3 is that a larger page size is used. This is in order to keep up with the many sessions that may simultaneously consume this entropy reserve, without frequent interruptions for unbuffered disk I/O to complete as pages are turned.

Another distinguishing feature of pad 00000 in Figure 3 is that it appears to be out of pages, as the receive page number is the same as its page count. In point of fact, this pad is not used for transmitting or receiving; it's the baggage *being* transmitted. The transmit page and offset entries are used to track this pad's consumption, and the receive page number is set such that various integrity checks that apply to the other pads are not violated.

The size of pad 00000 exactly equals the total amount of entropy that the endpoint sessions can receive. The five clients each have 128 MiB of key material available; that's 640 MiB. That means that at most, they can download 640 MiB of new key material. But since new keys must be distributed identically between *pairs* of endpoints, only 320 MiB of unique new key material is necessary. So when installing an OTP client key on a hub machine, additional entropy in the amount of half the client key's length are added to the reserve. This is how pad 00000 comes to have 80 pages at 4 MiB each, for a total of 320 MiB of fresh entropy for future distribution.

In section 3.4.5, it was divulged that I unsealed 1,024 MiB of pregenerated entropy for this project. Figure 3 shows where it was consumed, as 960 MiB is accounted for there. It also tells you that I didn't have enough entropy left in batch K8CLZN to add a sixth client with the same capacity as the others.

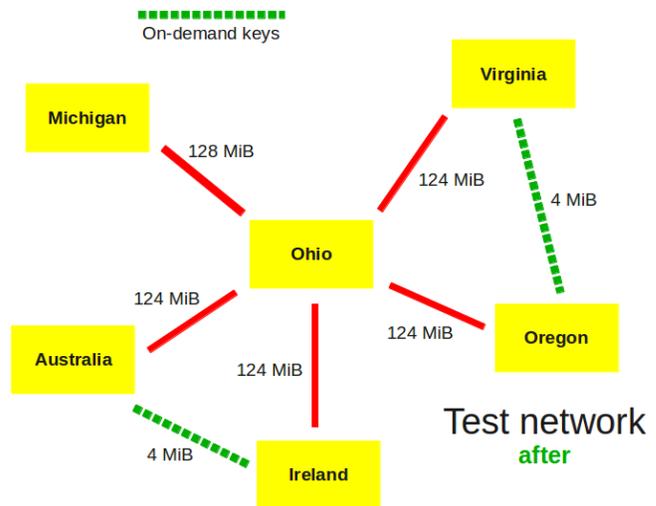

**Figure 8: Test network after conveying OTP material**

The time came to distribute some of this entropy reserve securely across the network. As mentioned in sections 3.4.3 and 3.4.4, this process wasn't *totally* secure, but the application and handling of the OTP itself was. It's like having a locked gate at the entrance of a New Mexico ranch; you can just walk around the gate. I manually removed — stole, if you will — pages 78 and 79 from pad 00000, updated `session.metadata` to indicate we never had those pages, and transmitted them via *iid*'s file transfer command to Virginia and Oregon (to share page 78), and Ireland and Australia (to share page 79). I can't remember what I did wrong, but I somehow mishandled these pages and leaked them into some RAM, to a disk, or both, so I discarded those and transferred pages 76 and 77 in order to get everything correct. One benefit of this retest is that I got a second set of elapsed time measurements. Figure 8 shows the configuration of the network after pages 78 and 79 were transferred.

To the best of my knowledge this was the first occasion in human history that a one-time pad was distributed securely across an open network. The most similar previous work appears in the field of quantum key distribution, which actually generates the same OTP at both ends, as opposed to distributing an arbitrary pre-existing OTP. My work is superior to quantum key distribution in the sense that it does not have QKD's expectation of a contiguous optical connection between endpoints. Furthermore, as the machine located in Sydney, Australia is more than 15,000 km from Ireland, my transmission beat the known QKD distance record of 148.7 km by a factor of 100.

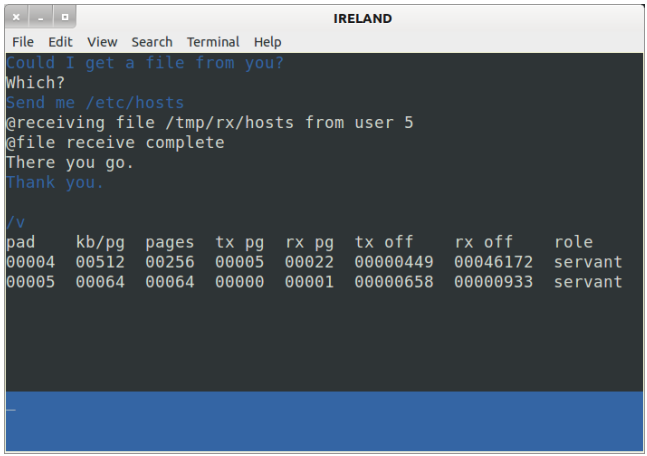

**Figure 9: Irish endpoint of a 15,000 km OTP conversation**

Figures 9 and 10 show brief chat dialogs across the newly-distributed Ireland-Australia one-time pad. An exchange involving a few words and a short file are shown. Also shown is output of the /v (display vault) command for both sessions. The reader can ascertain from the pad sizes and Figure 8 which pads connect with the hub, versus which connect to a client endpoint.

### 4.2) Speed measurements

A few packets were tossed around the test network to determine the system's responsiveness. It was known in advance that performance would be less than optimal because of the code's simplicity; every packet waits for its predecessor to be acknowledged by the receiving station. Bearing this in mind, here are the measurements.

Figure 11 shows that if the number of sessions is small, throughput is not affected by the number of sessions transmitting. This is an expected result, as the physical network and system resources are not in heavy use. The command to *iid* at the Ohio location was /g65536 in each instance; that is, transmit 65,536 bytes of ciphertext across the link. These were run three times from Ohio to each destination separately; e.g., Ohio to Michigan, get the result,

Ohio to Oregon, get the result, and so on. The three transmission times and their average in seconds are reported in the left side of the table.

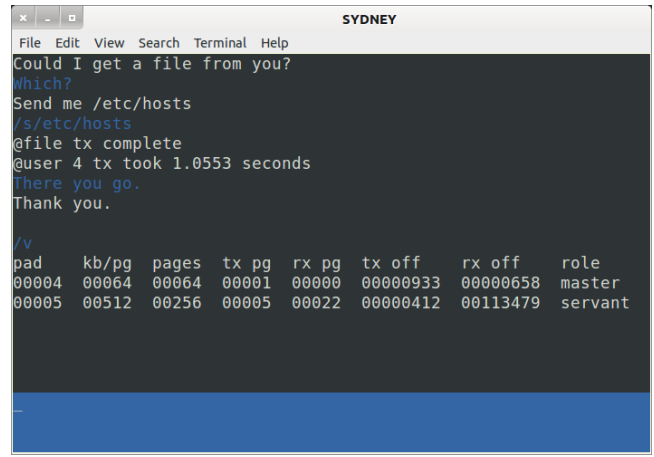

**Figure 10: A file is sent from Sydney, Australia to Ireland**

The test was repeated with the transmissions to all five destinations started simultaneously using the /b (as in batch) command. Transmission times for three trials of this "start five at once" test and their averages are shown on the right side of the figure.[21]

| | 64 KiB separately | | | | 64 KiB all at once | | | |
|---|---|---|---|---|---|---|---|---|
| to | #1 | #2 | #3 | avg | #1 | #2 | #3 | avg |
| MI | 2.1 | 2.1 | 4.4 | 3.1 | 3.1 | 3.2 | 3.0 | 3.1 |
| VA | 2.8 | 2.8 | 2.8 | 2.8 | 3.1 | 3.2 | 3.3 | 3.2 |
| OR | 5.2 | 5.2 | 5.2 | 5.2 | 5.7 | 5.8 | 5.7 | 5.7 |
| IE | 7.3 | 8.6 | 7.3 | 7.6 | 7.6 | 7.7 | 7.6 | 7.6 |
| AU | 13.6 | 12.4 | 16.6 | 14.2 | 14.1 | 15.2 | 15.2 | 14.8 |

**Figure 11: Short transmission times (in s) from Ohio**

Although the Figure 11 numbers might show a portion of the effect of using more than one session at a time, the numbers reported skew downward in the case of Oregon, Ireland, and Australia. This is because additional throughput became available as closer destinations finished their transfers. In any event, the results indicate that the transmissions are scaling well; no data transmission from the right hand set was hindered by the fact that the same hub was servicing four similar transmissions.

Elapsed times for the two mass conveyances of 4 MiB of key material from Ohio to the Virginia, Oregon, Ireland, and Australia clients appear in Figure 12. The transfers to these four destinations were initiated simultaneously for both attempts. Although not shown in any table, I found during setup that very few packets were being lost across any link; therefore the large differences in throughput stem from round trip time primarily and not from lost packets.

---

21  Although these 30 transmissions repeated the same plaintext, they used separate key material and had distinct ciphertexts.

|  | 4 MiB all at once | | |
| --- | --- | --- | --- |
| to | #1 | #2 | avg |
| VA | 185 | 195 | 190 |
| OR | 347 | 346 | 347 |
| IE | 467 | 481 | 474 |
| AU | 765 | 987 | 876 |

**Figure 12: Long transmission times (in s) from Ohio**

Once the new "on-demand" OTPs were established at their four endpoints, throughput tests were conducted across these links in both directions by making three trials each of `/g65536`. Results appear in Figure 13 and are staggeringly consistent, and I believe this is because traffic was routed through the hosting company's private network instead of the public Internet.

|  | 64 KiB separately | | |
| --- | --- | --- | --- |
| route | #1 | #2 | #3 |
| VA to OR | 4.7 | 4.7 | 4.7 |
| OR to VA | 4.7 | 4.7 | 4.7 |
| IE to AU | 16.6 | 16.6 | 16.6 |
| AU to IE | 16.6 | 16.6 | 16.6 |

**Figure 13: 64 KiB transmission times via new OTPs**

### 4.3) Effects of jamming

As mentioned in section 3.2, the current implementation incurs preventable inefficiencies as it attempts to match incoming cyphertext to a specific OTP at a possibly mis-synchronized offset. To quantify this problem, a short program was written to inject jamming packets into the network in order to make *iid* work harder.

The first trials at jamming resulted in a very high percentage of dropped packets. At that time, the run loop did not use the `select` call to multiplex input between the keyboard and network; instead, non-blocking reads were used with short naps taken in between. The interference was so great that the time I would have spent presenting numeric results was spent instead modifying the run loop to use `select`.[22] The fixed system works much better, although some limitations remain.

Jamming packets were generated on the Ohio machine, as it was the nearest point to the victim in the network. The victim *iid* ran on a different core of the same machine. As *iid* discards any packet shorter than the 16 byte header used in the OTP protocol, the jammer gained negligible advantage sending packets smaller than 16 bytes. Trials of various packet lengths showed that jamming's effectiveness per unit bandwidth was greatest when the malicious packets were 16 bytes long.[23] The jamming packets did not require any special content to succeed; although their content was read in advance from `/dev/urandom`, sending all zeros would have produced identical measurements.

| jamming frequency | jamming bit rate | CPU use | 100 kB tx time | receiver throughput |
| --- | --- | --- | --- | --- |
| none | 0 kbit/s | 0% | 2.1 s | 380 kbit/s |
| 390 Hz | 50 kbit/s | 66% | 2.2 s | 360 kbit/s |
| 760 Hz | 90 kbit/s | 100% | 67.2 s | 12 kbit/s |

**Figure 14: DoS attack by injecting random packets**

Although the victim machine was set up with only two OTPs to keep track of, a jamming rate below even DSL speeds caused throughput from Michigan to Ohio to fall almost 97%. Figure 14 has the numbers.[24] The bottleneck was that brute force attempts to find a pad and offset to decrypt the interfering packets successfully was causing the CPU to fall behind when reading incoming packets. Many packets were dropped as a result, necessitating delays and retransmissions.

### 5) Future work

The chief weakness of this research is that it has yet to produce a practical system for evaluation and use in the field.[25] In pursuit of something better, three interconnected classes of further work are advanced.

---

22 The jamming experiments were the first measurements taken, so the `select` issue was resolved prior to Figures 11, 12, and 13.

23 Implying that all these packets had a plaintext length of zero.

24 Unacknowledged packets were resent from Michigan after 0.3 seconds and every two seconds thereafter until acknowledged.

25 Notwithstanding, source code for *iid* is available on request under the terms of the GNU General Public License, Version 3.

## 5.1) Fixes

A first step in correcting this is to make several usability improvements to *iid* so that it works consistently. Another need is to incorporate IPv6. One motivation for IPv6 is that many network operators will require it, but in certain jurisdictions such as the United States, there is also a legal consideration.

In these jurisdictions, access to email under the law depends on the location where it is stored. The law affords email that is physically stored within the limits of a person's home much greater protection than than email that is in another location, such as on a server owned by another party.[26] The law reasons that when you store information on a computer that you do not own, you have voluntarily "disclosed" this information to the owner of the equipment you are using, and that your action has forfeited your right to privacy. Accordingly, the owner or possessor of the machine or media where your data are located can be compelled to disclose them to various authorities by means of subpoena or other lawful process.[27] In contrast, information stored within the boundaries of a private residence requires a search warrant to obtain.[28]

IPv6 will change the playing field by providing enough network addresses for users to dispense with third-party email and voice mail providers altogether. Instead, email will be passed directly from its sender's device to its receiver's device, affording legal protection against subpoenas as well as requirement of a search warrant prior to any seizure. This move away from central providers also will reduce the number of convenient collection points for illicit mass surveillance. In light of all this, IPv6 will be the protocol of choice for sensitive data in the future, and OTP implementations will have to support it.

Several performance enhancements to the software and ciphertext protocol are needed. The receiving end needs to be able to locate the correct pad and offset quickly, without any protocol changes which would make synchronization data available to a hostile party. Achieving this will incur a small per-packet overhead in bandwidth and entropy consumed, but the outcome will be a system that can be scaled to handle thousands or even millions of OTPs efficiently.

Fuller utilities are needed for the generation, transport, and installation of OTP material. My installation script did the hub's end of the job well enough, but it left the client with unresolved data remanence troubles.

Theft of key material from — or injection of key material into — an OTP hub will be an appealing attack vector unless some form of fall-back cryptography (block cipher, public key, etc.) is incorporated. This should provide a useful countermeasure in some instances until advances in cryptanalysis discredit the fallback cipher.

The protocol needs changed in order to consider the connection's bandwidth-delay product. This involves allowing a number of packets to be sent without intervening acknowledgements. Once this is working, addition of congestion control similar to TCP's may be needed for certain uses.

A reasonable (not manual) process for automatically distributing OTPs from a hub to its endpoints needs to be built in. This will require several decisions concerning the amount of key and network resources to expend in exchange for reduced latency and improved availability.

## 5.2) Interoperability

A generalized, interoperable model needs to be produced for broader testing and trial use. This won't be a split-screen terminal chat program.

I suggest that the best point of attachment to today's system and network infrastructure will be to implement a robust OTP model as a cipher suite for a mainstream TLS implementation.[29] It may turn out that two ciphers suites are necessary, with one being for conventional pre-shared keys and the other for hub-facilitated OTP networks.

The benefit of supporting TLS is that one can go anywhere from there in terms of applications and functionality. The outcome of this project could secure login sessions, email hops, video conferences, web applications, file transfers, authentication, and so on — the sky's the limit. A concern, however, is that configurations need to be set up and validated carefully, or the implementation might downgrade to a non-OTP cipher without alerting the administrator or end user.

## 5.3) People

Documentation, education, advocacy, and correspondence will be necessary to assist the public in the adoption,

---

26 This is grossly oversimplified; see for instance [14].

27 The national security letters of the United States are one example.

28 Once again for every rule, there is an exception. [15] gives a recent and thorough explanation to persons within the United States.

---

29 An in-depth comparison between implementations appears at [16].

testing, and refinement of the OTP products which emerge from this work. Involvement with standards organizations such as the Internet Engineering Task Force (IETF) will be necessary to support interoperability among OTP equipment systems and providers, and to reduce uncertainties faced by potential consumers.

As the export of cryptographic equipment and software is regulated in my jurisdiction, some compliance work may be necessary.

### 6) Closing thoughts

Ten years ago, block cipher expert Bruce Schneier blogged [11] that one-time pads "are useless for all but very specialized applications, primarily historical and non-computer." This statement is as false today as it was in 2002, but the complete text is, like much of Schneier's writing, highly entertaining and makes for enjoyable reading.

Some may argue that my hub distribution model is irredeemably flawed. They might insinuate that people who need one-time pads to communicate will either be "anti-social" and therefore not trust any hub operator, or they are "naive" and will hire their attacker to run the hub. I disagree for four reasons:

1.  The parties will not use hub-supplied entropy as received; there will be some further encryption.

2.  People who communicate in networks do so because they have some common interest at stake, such as a company they work for, church worship in, or family they vacation with.

3.  Communicants ordinarily have some idea who they don't trust or want to hear their exchanges. These participants need only agree on a facilitator they can trust more than the attacker they are most concerned about. Many times, that is not a very high bar to jump over.

4.  In a large network with millions of potential pairs of communicants, few people know each other well, but the party at the hub may be recognizable by all. The risk of a corrupt hub operator, who may be under a lot of scrutiny, may not be as great as the general risk of untrustworthy persons at various endpoints.

Quantum key distribution compares unfavorably against one-time pad approaches, even if its high cost of hardware,

cable, and installation is disregarded. It suffers even worse endpoint scalability problems than OTPs have. Revisiting the 10,000 user network problem one final time, the QKD approach involves laying 50,005,000 separate fiber paths.

Additionally, QKD has a very serious authentication problem in that it is only *secure* if the entire fiber run is in plain sight at the time the key is exchanged. Try that in a Vienna sewer when you want to make a bank transfer. In contrast, one-time pad media exchanges are easily witnessed by appointment, even for transoceanic exchanges.

To finish, one-time pads have five essential properties. They are practical. They are scalable. They are inexpensive. Their mathematical properties are provable. And last of all, although one-time pads have never been convenient, they indeed are as indispensable today as they proved to be in the 20th century.

### Credits

I cannot express enough gratitude for the enormous support of so many who contributed to my work's success. I thank and acknowledge each of you profusely and publicly, and I celebrate your shining contribution to freedom of thought, freedom of association, and freedom of participation by here withholding every last one of your names.

## Appendix A

**The *Eternity* production system for large OTPs**

As I need to earn 87 more credit hours to receive my degree, this Appendix is reserved for future study.